\documentclass[12pt,letterpaper]{article}
\pdfoutput=1

\usepackage[dvips]{epsfig}
\usepackage{color}
\definecolor{refkey}{rgb}{1,0,0}
\definecolor{labelkey}{rgb}{0,0,1}
\usepackage{amsfonts,amssymb,amsmath}

\newcommand{\be}{\begin{equation}}
\newcommand{\ee}{\end{equation}}
\newcommand{\ben}{\begin{displaymath}}
\newcommand{\een}{\end{displaymath}}
\newcommand{\bea}{\begin{eqnarray}}
\newcommand{\eea}{\end{eqnarray}}
\newcommand{\bean}{\begin{eqnarray*}}
\newcommand{\eean}{\end{eqnarray*}}

\def\g {\gamma}

\def\d {\delta}

\newcommand{\ads}[1]{\mbox{${AdS}_{#1}$}}
\newcommand{\adss}[2]{\mbox{$AdS_{#1}\times {S}^{#2}$}}

\newcommand{\eg}{{\it e.g.}}

\newcommand{\commentout}[1]{}

\newcommand{\beq}{\begin{equation}}
\newcommand{\eeq}{\end{equation}}
\newcommand{\beqr}{\begin{displaymath}}
\newcommand{\eeqr}{\end{displaymath}}
\newcommand{\beqa}{\begin{eqnarray}}
\newcommand{\eeqa}{\end{eqnarray}}
\newcommand{\beqar}{\begin{eqnarray*}}
\newcommand{\eeqar}{\end{eqnarray*}}

\newcommand{\n}{\nu}

\newcommand{\cN}{{\cal N}}

\newcommand{\cO}{{\cal O}}

\newcommand{\cA}{{\cal A}}

\newcommand{\half}{\ensuremath{\frac{1}{2}}}

\newcommand{\bz}{\ensuremath{\bar{z}}}

\newcommand{\N}[1]{\ensuremath{\cN=#1}}

\begin{document}

\title{\LARGE \bf Minimal area surfaces dual to Wilson loops and the Mathieu equation }

\author{
	 Changyu Huang\thanks{E-mail: \texttt{cyhuang@purdue.edu}} ,
	 Yifei He\thanks{E-mail: \texttt{he163@purdue.edu}} ,
	Martin Kruczenski\thanks{E-mail: \texttt{markru@purdue.edu}} \\
	Department of Physics and Astronomy, Purdue University,  \\
	525 Northwestern Avenue, W. Lafayette, IN 47907-2036.}

\maketitle

\begin{abstract}
 The AdS/CFT correspondence relates Wilson loops in \N{4} SYM to minimal area surfaces in \adss{5}{5} space. Recently, a new approach to study minimal area surfaces in $\ads{3}\subset\ads{5}$ was discussed based on a Schroedinger equation with a periodic potential determined by the Schwarzian derivative of the shape of the Wilson loop. Here we use the Mathieu equation, a standard example of 
a periodic potential, to obtain a class of Wilson loops such that the area of the dual minimal area surface can be computed analytically in terms of eigenvalues of such equation. As opposed to previous examples, these minimal surfaces have an umbilical point (where the principal curvatures are equal) and are invariant under $\lambda$-deformations. In various limits they reduce to the single and multiple wound circular Wilson loop and to the regular light-like polygons studied by Alday and Maldacena. In this last limit, the periodic potential becomes a series of deep wells each related to a light-like segment. Small corrections are described by a tight--binding approximation. In the circular limit they are well approximated by an expansion developed by A.Dekel. In the particular case of no umbilical points they reduce to a previous solution proposed by J. Toledo. The construction works both in Euclidean and Minkowski signature of \ads{3}.
\end{abstract}

\clearpage
\newpage



\section{Introduction}
\label{intro}
 The AdS/CFT correspondence \cite{malda} is a duality between gauge theories and string theory. In the large N limit \cite{largeN} and at strong 't Hooft coupling the strings can be treated semi-classically. Since the classical solution of the string equations of motion is given by an extremal surface, the study of minimal area surfaces in \ads{} space is crucial in exploring the correspondence. In particular  the relation between Wilson loops and minimal area surfaces ending at the boundary \cite{MRY} is an intense area of research \cite{cWL,WLref,WLint}. In the case of single contour, smooth Wilson loops that interests us in this paper, analytical results for the minimal area problem are known in the case of near circular Wilson loop \cite{SY,CHMS,Cagnazzo,Dekel} based on perturbation theory, and for the case
of surfaces with no umbilical points\footnote{An umbilical point is a point where both principal curvatures coincide, in terms of the Pohlmeyer reduction described below it is a zero of the analytic function $f(z)$.} \cite{IKZ,KZ,IK} using Riemann theta functions along the lines of \cite{BB,BBook} and a solution proposed by Toledo \cite{Toledo} using the Y-function method. For surfaces with umbilical points the only analytical results are for light-like Wilson loops with cusps \cite{cusp,scatampl,AMSV}. An exception us the circular Wilson loop since the dual surface is a half-sphere and all points are umbillical (both curvatures equal).  Finally let's point out that these developments are related to integrability properties of the sigma model that also appear in the closed string case \cite{ClosedStrings}.

 Going back to the minimal area problem, in this paper we concentrate in the case of Euclidean surfaces embedded in $EAdS_3$ and $AdS_3$ considered as subspaces of \adss{5}{5}. They are dual to space-like Wilson loops embedded in a subspace $\mathbb{R}^2\subset \mathbb{R}^{3,1}$ or $\mathbb{R}^{1,1}\subset \mathbb{R}^{3,1}$. 
For concreteness, in the rest of the introduction we review the $EAdS_3$ case, the $AdS_3$ case is similar. 
 A standard approach to the minimal area surface problem is to use the Pohlmeyer reduction \cite{Pohlmeyer}. First, the world-sheet is taken to be a unit disk and the induced metric is written in conformal coordinates $z=\sigma+i\tau$, namely in coordinates where it is conformally flat:
 \beq
 ds^2 = 4 e^{2\alpha} dz d\bz\, .
 \label{a1}
 \eeq
 for some function $\alpha(z,\bz)\in \mathbb{R}$. The crucial observation is that $\alpha(z,\bz)$ satisfies a generalized cosh-Gordon equation 
\beq
 \partial \bar{\partial} \alpha = e^{2\alpha} + f(z) \bar{f}(\bar{z}) e^{-2\alpha}\, ,
\label{a2}
\eeq
where $f(z)$ is a holomorphic function that depends on the shape of the Wilson loop. After solving this equation one can compute the regularized area by the formula
\beq
 \cA_f = -2\pi - 4 \int_D f\bar{f} e^{-2\alpha} d\sigma d\tau\, ,
\label{a3}
\eeq
where $D$ represents the unit disk $|z|\le 1$. This regularized area determines the expectation value of the corresponding Wilson loop in the field thepry at large 't Hooft coupling. 
When $f(z)$ has no zeros in the world-sheet, analytical solutions to the cosh-Gordon equation can be found using Riemann theta functions \cite{IKZ,KZ,IK}. 
Alternatively, given a nowhere vanishing $f(z)$ one can solve a Y-system type of equation \cite{Toledo} and from there compute the area. In the case where $f(z)$ has zeros, the only known analytical results are for the Minkowski case and then only for Wilson loops with light-like boundaries \cite{AMSV}. 

In this paper we find analytical results for the case where $f(z)=f_0 z^n$, namely $f(z)$ has a multiple zero and the Wilson loop is given by a smooth space-like curve with no cusps. In order to do that, we use a way to approach the problem described in \cite{WLMA} where it was argued that such surfaces can be studied by considering a Schroedinger equation defined on the Wilson loop. Indeed, consider for concreteness the case of the Poincar\'e patch of $EAdS_3$ whose boundary is $\mathbb{R}^2$ and suppose that such Wilson loop is given by an arbitrary parameterization 
 $X(s) = X_1(s)+i X_2(s)$, $s=0\ldots 2\pi$. Then one can define a Schroedinger equation
\beq
 -\partial_s^2 \chi + V(s) \chi(s) =0\, ,
\label{a4}
\eeq
 where $V(s)$ is a complex potential given by 
\beq
 V(s) = V_0(s) + i V_1(s) =-\half  \{X(s),s\} \, ,
\label{a5}
\eeq
where $\{X(s),s\}$ denotes the Schwarzian derivative.
 Up to global conformal transformations we can reconstruct the shape of the Wilson loop by considering two linearly independent solutions of such equation and taking the ratio
\beq
 X(s) = \frac{\chi_1(s)}{\chi_2(s)}\, .
\label{a6}
\eeq
 Such solutions are anti-periodic as can be seen from their explicit form:
\beq
\chi_1(s) = \frac{X(s)}{\sqrt{\partial_s X(s)}}, \ \ \ \chi_2 = \frac{1}{\sqrt{\partial_s X(s)}}\, .
\label{a7}
\eeq
 The method proceeds by defining a generalized potential depending on a complex spectral parameter $\lambda$:
\beq
 V(\lambda,s) = V_0(s) 
  + \frac{i}{2} \left(\lambda+\frac{1}{\lambda}\right) V_1(s) 
  + \half       \left(\lambda-\frac{1}{\lambda}\right) V_2(s)  \, ,
\label{a8}
\eeq 
where $V_2(s)$ is such that the solutions of the corresponding Schroedinger equation are anti-periodic for any value of the parameter $\lambda$. Since the solutions of a Schroedinger equation with periodic potential are usually quasi-periodic, this provides an infinite set of condition that should be sufficient to determine the function $V_2(s)$. 
 Once $V_2(s)$ is found the area follows from a simple formula. In this paper we used this method to find new Wilson loops whose shape and area can be computed analytically. The shape of the dual surface can then be obtained numerically.

\section{Euclidean \ads{3} case}

In this case we use Poincar\'e coordinates $(X_1,X_2,Z)$ where $EAdS_3$ has a metric
\beq
 ds^2 = \frac{dX_1^2 + dX_2^2 + dZ^2}{Z^2}\, .
\label{b1}
\eeq
The boundary is at $Z=0$ and the Wilson loop is given by a curve
\beq
X(s) = X_1(s) + i X_2(s)\, .
\label{b2}
\eeq
where $X$ is a complex coordinate. The world--sheet is parameterized by a complex coordinate $z=re^{i\theta}$, $|z|=r\le 1$.  The boundary conditions for the surface are that $Z(e^{i\theta})=0$ and $X(e^{i\theta})=X(s(\theta))$ for some reparameterization $s(\theta)$. 

 The method to find minimal area surfaces that we proposed in \cite{WLMA} is based on obtaining anti-periodic solutions of a Schroedinger equation with periodic potential. The prototypical periodic Shroedinger equation is the well-known Mathieu equation \cite{Mathieu}: 
\beq
 \partial_u^2 \chi(u) + (a-2q\cos 2u) \chi(u) =0 \, ,
\label{b3}
\eeq
where $q$ is a parameter and $a$ plays the role of energy eigenvalue. It has quasi-periodic solutions known as Floquet solutions 
\beq
 \chi_\nu(u+\pi) = e^{i\pi\nu} \chi_\nu(u)\, ,
\label{b4}
\eeq
 where the Mathieu characteristic $\nu=\nu(a,q)$ is a function of the parameters $a$ and $q$. For given $q$ we can fix $\nu$ to any desired value by choosing an appropriate value of $a=a_\nu(q)$. 
  It is therefore natural to look for minimal area surfaces based on the Mathieu equation. In order to do that, we write the potential in eq.(\ref{a8}) as
\beq
 V(\lambda,s) = -\frac{1}{4} + 6 \beta_2 - \lambda f_0 e^{i(n+2) s} + f_0 \frac{1}{\lambda}  e^{-i(n+2) s}\, ,
\label{b5}
\eeq
corresponding to the real and periodic functions
\beq
 V_0(s) = -\frac{1}{4} + 6\beta_2, \ \ \ V_1(s) = -2f_0\sin(n+2)s, \ \ \ V_2(s) = -2f_0\cos(n+2)s\, .
 \label{b5a}
\eeq
 The independent parameters are $n$, an even integer and $f_0$, a real constant. The constant $\beta_2$, however, is not arbitrary. The reason is that, if this potential corresponds to solutions of the minimal area problem, then the solutions to the corresponding Schroedinger equation should be anti-periodic for any value of $\lambda$.  Consider the case when $\lambda=e^{i\varphi}$. It is clear that the potential is the same up to a shift in the variable $s$ and therefore the periodicity of the solutions will be the same for any value of $\varphi$ and, by analyticity, for any complex value of $\lambda$. Therefore we only need to ensure that the solutions are anti-periodic for $\lambda=1$ which is easily done by choosing $\beta_2$ appropriately. Indeed, by defining
\beq 
u(s)=\frac{(n+2)s+\varphi}{2}+\frac{\pi}{4}\, ,
\label{b6}
\eeq
the equation
\beq
 -\partial_s^2 \chi(s) + V(\lambda,s) \chi(s) = 0\, ,
\label{b7}
\eeq
becomes the standard Mathieu equation with
\beq a=\frac{1-24\beta_2}{(n+2)^2},\ q=\frac{4 i f_0}{(n+2)^2}\, .
\label{b8}
\eeq
Now we look for a Floquet solution of Mathieu equation with real characteristic exponent $\n$, i.e.
\beq
\chi_\n(u+\pi)=e^{\pi i\n}\chi_\n(u)\, ,
\label{b9} 
\eeq
or equivalently
\beq
 \chi_\n(u) = e^{i\n u}\, p_\nu(q,u)\, ,
\label{b10}
\eeq
where $p_\n(\theta)$ is $\pi$ periodic. 
Since we require $\chi(s)$ to be an anti-periodic function of $s=0\ldots 2\pi$, it follows that, when $\Delta s=2\pi$, 
\beq
\begin{aligned}
     \Delta u &= (n+2) \pi\, , \\ 
	(2k+1)\pi &= \n \Delta u\, ,
	\Rightarrow\  &\n=\frac{2k+1}{n+2}\ \ \ (k\in \mathbb{Z})\, ,
\end{aligned}
\label{b11}
\eeq
 Thus there is a discrete family of solutions labeled by $k\in\mathbb{Z}$. Such solutions, however do not necessarily have
 one boundary. Later we will see that for $k=0$ they do (and also, in certain cases, for $k=n+1$).  
 For each $\nu$ the Mathieu eigenvalue $a=a_\nu(q)$ can be determined\footnote{There are efficient numerical algorithms to obtain the eigenvalues, for example Mathematica implements such function $a_\nu(q)$.} and from there the constant 
\beq 
\beta_{2}=\frac{1}{24}\left[1-(n+2)^2a_\nu(q)\right]\, ,
\label{b12}
\eeq 
that, as seen in the next section, gives the area of the minimal surface. 
The shape of the Wilson loop is given by the ratio of two independent solutions of the Mathieu equation
\beq
X(u)=\frac{\chi_\n(u)}{\chi_{\n}(-u)} 
= e^{2i\n u}\,\frac{p_\n(q,u)}{p_\n(q,-u)} =\frac{Mc(u)+iMs(u)}{Mc(u)-iMs(u)}\, ,
\label{b13}
\eeq
where we used that replacing $u\rightarrow-u$ gives another solution of the Mathieu equation. For completeness, we also wrote the result in terms of the often used Mathieu sine and cosine defined as the odd and even solutions respectively.  All Mathieu functions are evaluated for $q$ given in eq.(\ref{b8}) and the eigenvalue $a=a_\nu(q)$. 

At this point it is useful to make contact with the Pohlmeyer reduction method. Indeed, from \cite{WLMA}, the potential is given by
\beq
 V(\theta) = -\frac{1}{4} + 6\beta_2(\theta) - f(\theta) \lambda e^{2i\theta}
+\frac{1}{\lambda} e^{-2i\theta} \bar{f}(\theta)\, ,
\label{b14}
\eeq
which means that, if we identify $s=\theta$, these solutions correspond to a holomorphic function 
\beq
 f(z) = f_0 z^n\, ,
\label{b15}
\eeq
which has no poles inside the disk, therefore justifying the choice $s=\theta$. 
The function $\beta_2(\theta)$ is constant and given by eq.\eqref{b12}. Replacing in the generalized cosh-Gordon equation (\ref{a2}), one can see that the equation is solved by a rotationally invariant function $\alpha(r)$ satisfying 
\beq
 \partial_r^2 \alpha(r) + \frac{1}{r} \partial_r \alpha(r) = 4 e^{2\alpha(r)} + 4 f_0^2 r^{2n} e^{-2\alpha(r)}\, .
\label{b16}
\eeq
Near the boundary $r\rightarrow 1$ this equation implies that, in terms of the variable $\xi=1-r^2$,
\beq
 \alpha = -\ln \xi + \beta_2 \xi^2 + \beta_2 \xi^3 + \cO(\xi^4), \ \ \ (\xi=1-r^2\rightarrow 0)\, ,
\label{b17}
\eeq
as already derived in \cite{WLMA} but with the observation that here $\beta_2$ is a constant independent of $\theta$.  
 In the Minkowski case, such function $f(z)= f_0 z^n$ was studied by Alday and Maldacena \cite{AM2}\footnote{They were called ``regular polygon'' solutions.} where it was noticed that eq.(\ref{b16}) is equivalent to the Painleve III equation. However only the case of an infinite world-sheet was considered, corresponding to a light-like Wilson loop with cusps. For the case of smooth Wilson, loops J. Toledo recently found one such example of solution using his Y-system method \cite{Toledo}. Here it corresponds to the case $n=0$ where $f(z)$ does not vanish anywhere on the world-sheet. On the other hand the solutions presented here correspond to the case of smooth Wilson loops where $f(z)$ has zeros, a case where no exact results for the area were known before. This equation was also studied in \cite{Novokshenov} in relation to minimal area surfaces but the surfaces considered there were different (multiple boundaries) and the area was not computed. Nevertheless, those results have some overlap with the Euclidean case considered in this paper. 

\section{Computation of the Area}

To compute the area we can use simple integration by parts in eq.(\ref{a3}). In order to do that we observe that eq.(\ref{b16}) implies that
\beq
 \partial _{r}\left[
   r^2(\partial _{r}\alpha)^2
   +2r\partial _{r} \alpha
   -4r^{2}e^{2\alpha}
   +4f_0^{2}r^{2n+2}e^{-2\alpha}
   \right] = 8(n+2)f_0^2r^{2n+1}\, e^{-2\alpha}\, .
\label{c1}
\eeq
 Thus
\beq
\begin{aligned}
 \cA_f + 2\pi &= -4 \int_D f\bar{f} e^{-2\alpha}\, r dr\, d\theta = -8 \pi f_0^2 \int_0^1 r^{2n+1} e^{-2\alpha(r)} \, dr \\
  &= - \frac{\pi}{n+2}\int_0^1
  \partial _{r}\left[
  r^2(\partial _{r}\alpha)^2
  +2r\partial _{r} \alpha
  -4r^{2}e^{2\alpha}
  +4f_0^{2}r^{2n+2}e^{-2\alpha}
  \right]\,dr \\
  &= \frac{-\pi}{(n+2)}
 \left. \left[
  r^2(\partial _{r}\alpha)^2
  +2r\partial_{r}\alpha
  -4r^{2}e^{2\alpha}
  \right]\right|_{r\rightarrow 1}\\
  &= \frac{24\pi\beta_2}{n+2}\, ,
\end{aligned}
\label{c2}
\eeq
where we used the behavior (\ref{b17}) for $\alpha$ at the world-sheet boundary. In this case $\beta_2<0$ implying that $\cA_f<-2\pi$. Finally, using eq.(\ref{b12}), the area can be put in terms of the eigenvalues $a_\nu(q)$ of the Mathieu equation 
\beq
 \cA_f = -2\pi +\frac{\pi}{n+2} - (n+2)\pi a_\nu(q)\, .
\label{c3}
\eeq
This concludes the computation of the area. However it is important to point out that the fact that the integrand in (\ref{a3}) is a total derivative is not a coincidence but quite generic. Indeed, in \cite{WLMA} and \cite{IK3}, using integration by parts, a formula for the area in terms for the Schwarzian derivative of the contour was given. It is valid whenever $f(z)$ has no zeros on the world-sheet, clearly not the case here since
 $f(z) = f_0 z^n$. The problem in using the formula is that, when $f(z)$ has zeros, the function $\sqrt{f}$ has cuts and the integration by parts in eq.(111) of \cite{IK3} gives rise to integrals around the cuts as in \cite{AMSV}.  However, in the present case, when $n$ is even the function  $\sqrt{f}$ is well--defined, thus we can define the function
 \beq
  W(z) = \int_0^z \sqrt{f(z')} dz' = \frac{2\sqrt{f_0}}{n+2}\,  z^{\frac{n}{2}+1} \, ,
 \label{c4}
 \eeq
 and use the same argument as in \cite{IK3}. It starts with the observation that the generalized cosh-Gordon equation implies 
\beq
\begin{aligned}
j &= j_z dz + j_{\bz} d\bz \, ,\\
j_z &= 4f\sqrt{\bar{f}} e^{-2\alpha} \, ,\\
j_{\bz} &= \frac{2}{\sqrt{\bar{f}}}\left[\bar{\partial}^2\alpha-(\bar{\partial}\alpha)^2\right] \, ,\\
dj &= 0\, ,
\end{aligned}
\eeq
and therefore
\beq 
\begin{aligned}
\cA_f + 2\pi &= - 4 \int_D f\bar{f} e^{-2\alpha} d\sigma d\tau  \\
&= -\frac{i}{2} \int_D j\wedge d\bar{W} = \frac{i}{2}\int_D d(\bar{W}j) \\
&= \frac{i}{2}\oint_{\partial D} \bar{W} (j_z\,dz+j_{\bz}\,d\bz) \, .
\end{aligned}
\label{c5}
\eeq
 A caveat is that in \cite{IK3} the component $j_{\bz}$ was defined with an extra $\{\bar{W},\bz\}$ so that $j$ transforms properly as a one form. Here this is not possible since $\{W,z\}$ has a double pole at zero and therefore it would give an extra contribution to the contour integral.
Continuing the reasoning along the lines of \cite{IK3}, the boundary behavior of $j_{\bz}$ is given by     
\beq
 j_{\bz} = \frac{1}{\sqrt{\bar{f}}}(12 \beta_2(\theta) e^{2i\theta}) + \cO(\xi) \ \ \ \ \ (\xi=1-r^2\rightarrow 0)\, ,
\label{c6}
\eeq
 whereas $j_z\rightarrow 0$ and therefore
\beq
\cA_f = -2\pi + \frac{i}{2} \oint_D \frac{\bar{W}}{\sqrt{\bar{f}}} 12\beta_2 e^{2i\theta} \partial_\theta\bz \, d\theta\, .
\label{c7}
\eeq
Replacing $W(z)$ from eq.\eqref{c4}, $f=f_0 z^n$ and $z=e^{i\theta}$ it follows that
\beq
 \cA_f= -2\pi + \frac{24\pi\beta_2}{n+2}\, ,
\label{c8}
\eeq
in perfect agreement with eq.(\ref{c2}).

\section{Numerical checks, examples and limits of the solutions}

 The previous sections give analytic results for the shape of the Wilson loop (boundary curve) and for the area of the dual minimal area surface. It does not however give an analytic expression for the shape of the surface. In this section we use a numerical procedure to independently find these solutions (including the shape of the minimal surface) providing a numerical test of the previous results. In order to do so we first solve numerically the generalized cosh-Gordon equation for $\alpha$, eq.(\ref{a2}) and then the linear problem \cite{WLMA} for $\psi_{1,2}$ that leads to the shape of the surface and boundary contour. Finally, the integral in eq.(\ref{a3}) can be done numerically providing a check of the previous results.

\begin{figure}
 	\centerline{\includegraphics[width=\textwidth]{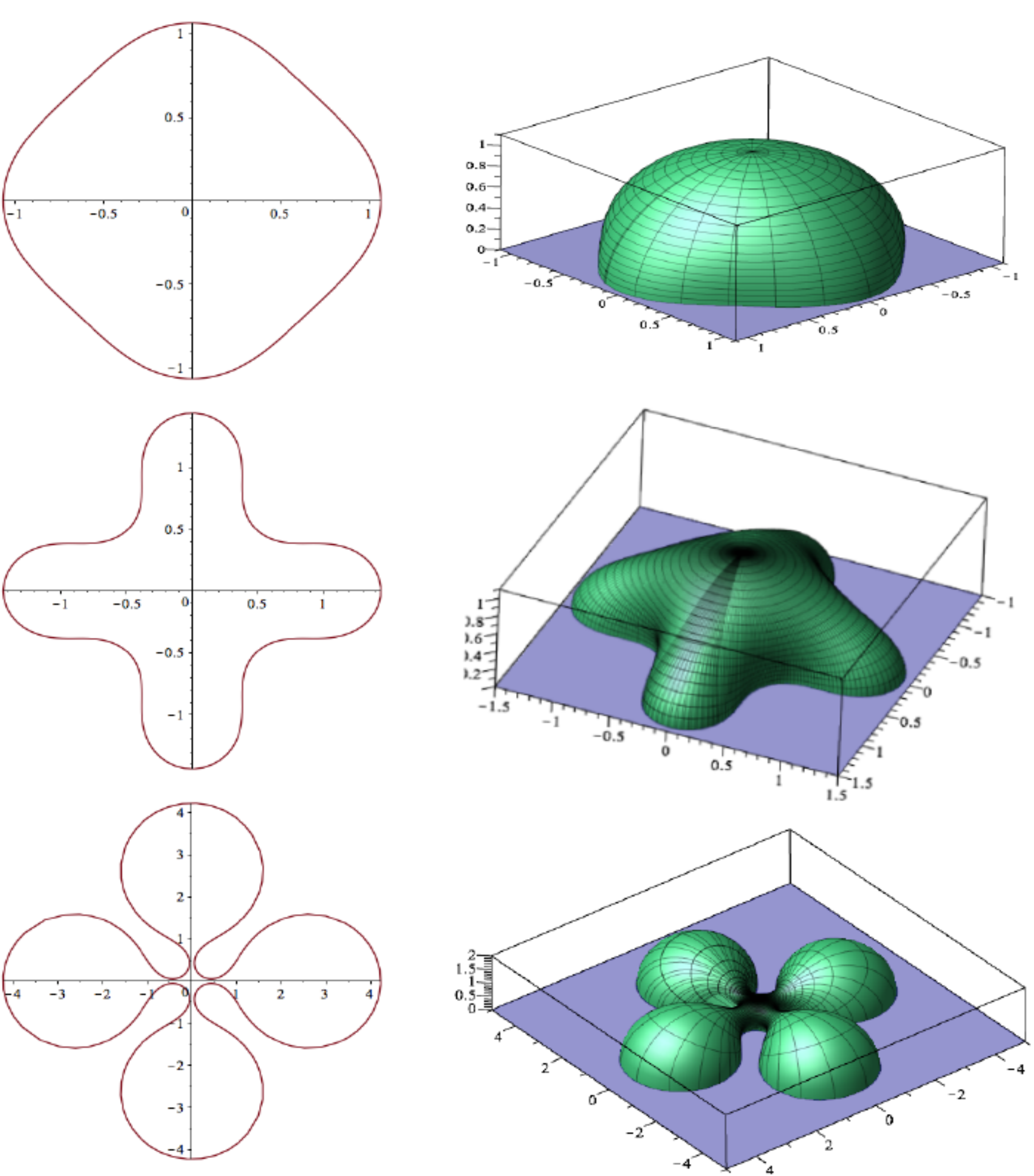}}
 	\caption{Surface corresponding to $n=2$ and the parameters $[\tilde{\alpha}_0, r_0, \beta_2, \cA_f]$ equal to (top to bottom): $[0.01,0.986067,-0.0200117, -6.660397]$, $[-.45, 1.4164114, -.43083877,-14.404305]$, $[-.95, 1.5079094, -1.3248754,-31.2565]$ }
 	\label{SurfacesE}
 \end{figure}

 With the choice $f=f_0z^n$, $f_0\in \mathbb{R}$, the generalized cosh-Gordon equation reads
\beq
 \partial \bar{\partial} \alpha = e^{2\alpha} + f_0^2 |z|^{2n} e^{-2\alpha} \, ,
\label{d1}
\eeq
and therefore has solutions that depend only on the radial coordinate $\alpha(r)$ where $z=r e^{i\theta}$. In this case the equation becomes
\beq
\frac{1}{4}\left[ \partial_r^2 \alpha + \frac{1}{r} \partial_r \alpha \right] =  e^{2\alpha} + f_0^2 r^{2n} e^{-2\alpha}\, .
\label{d2}
\eeq 
 For numerical purposes, it is convenient to define $r_0$ such that $f_0=r_0^{n+2}$ and then rescale $r= \frac{\tilde{r}}{r_0}$ and introduce $\tilde{\alpha}=\alpha-\ln r_0$ so that the equation becomes 
\beq
 \frac{1}{4}\left[ \partial_{\tilde{r}}^2 \tilde{\alpha} + \frac{1}{\tilde{r}} \partial_{\tilde{r}} \tilde{\alpha}\right] =  e^{2\tilde{\alpha}} +  \tilde{r}^{2n} e^{-2\tilde{\alpha}}\, .
\label{d3}
\eeq
Now we choose a value of $\tilde{\alpha}(\tilde{r}=0)=\tilde{\alpha}_0$ and, using the boundary condition $\left.\partial_{\tilde{r}}\tilde{\alpha}\right|_{\tilde{r}=0}=0$, integrate the differential equation up to a value $\tilde{r}=r_0$ where $\tilde{\alpha}$ diverges. This allows us to find $r_0(\tilde{\alpha}_0)$, namely for numerical purposes $\tilde{\alpha}_0$ is the parameter that defines the solution and $f_0$ is derived. To reconstruct the surface we now use the formulas (and notation) of \cite{WLMA}. First the linear problem 
\beqa
 \partial_{\tilde{r}} \psi_1 &=& \left(e^{i\theta+\tilde{\alpha}}-\tilde{r}^n e^{-\tilde{\alpha} - i (n+1) \theta}\right)\, \psi_2 \, , \\ 
 \partial_{\tilde{r}} \psi_2 &=&  \left(e^{-i\theta+\tilde{\alpha}}+\tilde{r}^n e^{-\tilde{\alpha}+i(n+1)\theta}\right)\, \psi_1\, ,
 \label{d3a}
\eeqa
is easily solved numerically\footnote{Numerically, eq.(\ref{d3}) and the linear problem are solved simultaneously.} and the shape of the surface reconstructed by first taking two linearly independent solutions $(\psi_1,\psi_2)$, $(\tilde{\psi}_1,\tilde{\psi}_2)$. For example by using the initial conditions, $(\psi_1(0)=1, \psi_2(0)=0)$ and $(\tilde{\psi}_1(0)=0, \tilde{\psi}_2=1)$. With those soltions, the following matrices can be constructed:
\beq
 \mathbb{A} = 
 \left(\begin{array}{cc} \psi_1 & \psi_2 \\ \tilde{\psi}_1 & \tilde{\psi}_2\end{array}\right), 
 \ \ \ \ \mathbb{X} = \mathbb{A}.\mathbb{A}^\dagger\, .
 \label{d3b}
\eeq
The Poincar\'e coordinates are identified, in terms of the components of $\mathbb{X}$ as
\beq
 Z = \frac{1}{\mathbb{X}_{22}}, \ \ \ \ X = X_1+iX_2 = \frac{\mathbb{X}_{21}}{\mathbb{X}_{22}}\, .
\label{d3c}
\eeq
 This method is rather crude and can be much improved by solving the equation using series expansions around $\tilde{r}=0$ and $\tilde{r}=r_0$ and then matching the expansions at some intermediate point to determine the function $r_0(\tilde{\alpha}_0)$. For example for $n=2$ the expansions read 
\beqa
 \tilde{\alpha} &=& \tilde{\alpha}_0 + e^{2\,\tilde{\alpha}_0} \tilde{r}^2 + \half  e^{4\,\tilde{\alpha}_0 } \tilde{r}^4+ \left( {\frac {{e^{6\,\tilde{\alpha}_0}}}{3}}+{\frac {{
  e^{-2\,\tilde{\alpha}_0}}}{9}} \right) \tilde{r}^6+  \ldots\, ,
  \\
 \tilde{\alpha} &=& -\ln r_0- \ln \xi +\beta_2\, \xi^2+\beta_2\, \xi^3 + \frac{1}{10} \left( r_0^{20}+2 \beta_2^2+9 \beta_2 \right) \xi^4+ \ldots\, ,
\label{d4}
\eeqa
where $\xi=1-\frac{\tilde{r}^2}{r_0^2}$. 
 The resulting shapes and values for the area match perfectly the results obtained from the Mathieu equation. To get an idea of the shape we present some typical results in figure \ref{SurfacesE}.

\subsection{Limits} 

Further, to get a better understanding of the solutions we consider several values of $n$ and swipe the values of $\tilde{\alpha}_0$ from minus to plus infinity. In the limit $\tilde{\alpha}_0\rightarrow \infty$ the exponential term $e^{-2\alpha}$ in the cosh-Gordon equation (\ref{d1}) becomes negligible and the solution approaches the one with $f(z)=0$, namely the circular Wilson loop. In the Mathieu equation $q\rightarrow 0$ and we can use standard techniques to approximate the solutions as deformations of the circle (see appendix). For finite values of $\tilde{\alpha}_0$ the Wilson loop resembles a smoothed out regular polygon with $n+2$ sides. As $\tilde{\alpha}_0$ decreases, the shape of the Wilson loop changes until at a certain value there is a discontinuity where the Mathieu characteristic $\nu(a,q)$ jumps from $\nu=\frac{1}{n+2}$ to $\nu=2-\frac{1}{n+2}$ and the Wilson loops becomes multiple wound. In the limit $\tilde{\alpha}_0\rightarrow -\infty$ it becomes a multiple wound circle, the winding number is $k=2n+3$.
The constant $\beta_2$ takes the value $\beta_2=-\frac{1}{6}(n+1)(n+2)$ and the area is $\cA_f=-2(2n+3)\pi$.

Finally it is also worth mentioning the limit $n\rightarrow \infty$ in which case the function $r^{2n}$ vanishes in the relevant region ($r<1$). In this case the solution approaches the circular Wilson loop.

\subsection{Relation to Painleve III equation}

We conclude this section by mentioning that, as noted in \cite{Novokshenov} for the Euclidean and \cite{AM2}for the Minkowski case, the radial cosh/sinh-Gordon equation is equivalent to the Painleve III equation. Indeed, the change of variables
\beq
\begin{aligned}
y &=\frac{1}{f_0}\frac{1}{r^n} e^{2\alpha}\, , \\
t &=r^{n+2}\, ,
\end{aligned}
\label{d5}
\eeq
reduces eq.(\ref{d2}) to the standard form
\beq
\frac{y^2}{t}\left(\frac{d}{d\ln t}\right)^2(\ln y)
=tyy''+yy'-t(y')^2
= b y + a y^3 +t(\delta + \gamma y^4)\, ,
\label{d6}
\eeq
with parameters
\beq
b=a=\frac{8f_0}{(n+2)^2}=-2iq, \ \ \d=\g=0\, .
\label{d7}
\eeq
The function $y(t)$ has a singularity at $t=1$ corresponding to $r=1$ namely the world-sheet boundary. Using the expansion (\ref{b17}) and after the change of variables (\ref{d5}) it follow that
\beq
y(t) =  \frac{i}{q}\frac{1}{(1-t)^2} + \frac{i}{12}\frac{1-4a_\nu(q)}{q} (1+(1-t)) + \cO[(1-t)^2]\, ,
\label{d8}
\eeq
namely, the solution $y(t)$ that has a double pole singularity at $t=1$ and the constant and linear coefficients in the Laurent expansion are determined by the eigenvalues $a_\nu(q)$ of the Mathieu equation. This provides an interesting relation between the Painleve III and Mathieu equation, namely that the Mathieu eigenvalues control the behavior near the movable singularity \cite{Novokshenov}. 

\section{Comparison with wavy-line approximation}

 In the limit $f_0\rightarrow 0$ the solutions approach the circular Wilson loop. In the Mathieu equation this corresponds to the limit $q\rightarrow 0$ that can be studied by standard perturbative methods. In this  section, however we employ the method developed by A. Dekel in \cite{Dekel} to study an arbitrary near circular Wilson loop providing then a check of that method. The idea is to assume that the contour has the shape given by the ratio of Mathieu functions in eq.(\ref{b13}), expand it for small $q$ and use Dekel's method to compute the area and check the result by comparing with eq.(\ref{c3}). We start by parameterizing the contour as a perturbation of a circular Wilson loop:
\begin{equation}
X(\theta)=e^{i\, s(\theta)+\xi(s(\theta))}\, .
\label{e1}
\end{equation}
Here $s(\theta)$ is the correct parametrization which we assume is unknown, and $\xi(s)$ to the third order of the perturbation takes the form (see appendix)
\begin{equation}
\xi(s)=a (-i q) \sin \left(\frac{s}{\nu }\right)+(-i q)^3 \left(b \sin \left(\frac{s}{\nu }\right)+c \sin \left(\frac{3 s}{\nu }\right)\right)\, ,
\label{e2}
\end{equation}
with
\begin{equation}
\begin{aligned}
& a=\frac{\nu }{\nu ^2-1},\\
& b=-\frac{3 \nu  \left(\nu ^2+5\right)}{16 \left(\nu ^2-4\right) \left(\nu ^2-1\right)^3}\, ,\\
& c=-\frac{\nu  \left(5 \nu ^4+36 \nu ^2-89\right)}{48 \left(\nu ^2-9\right) \left(\nu ^2-4\right) \left(\nu ^2-1\right)^3}\, .
\end{aligned}
\label{e3}
\end{equation}
If we let $\epsilon=-iq$, $p=\frac{1}{\nu}$, then the contour becomes
\begin{equation}
X(\theta)=e^{is(\theta)+\epsilon a \sin (p s(\theta ))+\epsilon^3 (b\sin (p s(\theta ))+c\sin (3p s(\theta ))) }\, .
\label{e4}
\end{equation}
With such parametrization, we can use the method introduced by Dekel in  \cite{Dekel} for the area calculation of the minimal surface ending on $X(\theta)$. Here we briefly review the procedure.

The expression for the regularized area is given in eq.\eqref{a3}. Therefore, 
we need to have $f(z)$ and solve the generalized cosh-Gordon equation for $\alpha(z,\bar{z})$ to calculate the area. These functions can be expanded as
\begin{equation}
\begin{aligned}
f(z)&=\sum _{n=1}^{\infty}f_n(z)\epsilon^n\, .\\
\alpha(z,\bar{z})&=\text{ln}(\frac{1}{1-z\bar{z}})+\sum _{n=2}^{\infty}\alpha_n(z,\bar{z})\epsilon^n\, .
\end{aligned}
\label{e6}
\end{equation}
Meanwhile, the correct parametrization $s(\theta)$ has the expansion
\begin{equation}
s(\theta)=s(\theta)=\theta+\sum _{n=1}^{\infty} s_n(\theta)\epsilon^n\, .
\label{e7}
\end{equation}
When $\epsilon=0$, $X(\theta)$, $s(\theta)$, $f(z)$ and $\alpha(z,\bar{z})$ reduce to the results of circular Wilson loop.
From the given boundary contour $X(s(\theta))$ we can first calculate the real and imaginary parts of the Schwarzian derivative expressed with the unknown $s_n(\theta)$. Based on the relations \cite{WLMA}
\begin{equation}
\begin{aligned}
&\text{Re}\{X(\theta), \theta\}=\frac{1}{2}-12\beta_2(\theta)\, ,\\
&\text{Im}\{X(\theta), \theta\}=-4\text{Im}(e^{2i\theta}f(\theta))\, ,
\end{aligned}
\label{e7a}
\end{equation}
we then expand the LHS of the equations with the parameter $\epsilon$ and extract $f(\theta)$ and $\beta_2(\theta)$ order by order. Next we plug $f(z)$ into the generalized cosh-Gordon equation to solve for $\alpha(z,\bar{z})$ and expand the solution around $r=1$ to get $\beta_2(\theta)$ which we use to compare with \eqref{e7a} to fix $s_n(\theta)$. In the end, we plug $s_n(\theta)$ back into $f(z)$, $\alpha(z,\bar{z})$ and \eqref{e5} to get the area.

As it turns out, the term $c\sin(3 p\,s(\theta))$ in the contour parametrization given in eq.\eqref{e2} appears in the area only at order $\epsilon^{6}$ and higher. Besides, to order $\epsilon^{4}$, the $b\sin(p\,s(\theta))$ term appears in the area expression as a first order effect, i.e., if we parametrize the contour as
\begin{equation}
X(\theta)=e^{is(\theta)+\epsilon a' \sin (p\,s(\theta ))+O(\epsilon^3)}\, ,
\label{e8}
\end{equation}
and let $a'=a+b\epsilon^2$, the area calculation will give the same result as the one given by \eqref{e2} to order $\epsilon^{4}$. Since we are comparing the area with the one given by Mathieu function only to the fourth order of the perturbation, we will adopt the parametrization \eqref{e8} and make the replacement
\begin{equation}
a'=\frac{\nu }{\nu ^2-1}+\frac{3 \nu  \left(\nu ^2+5\right) q^2}{16 \left(\nu ^2-4\right) \left(\nu ^2-1\right)^3}\, ,
\label{e9}
\end{equation}
at the end.

Our calculation shows that the expansions of the functions have the following form
\begin{equation}
\begin{split}
s(\theta)&=\theta-\frac{a'^2(p+5p^3)\text{sin}(2p\theta)}{8(-1+4p^2)} \epsilon^2+O(\epsilon^4)\, ,\\
f(z)&=-\frac{1}{4} i a' p \left(p^2-1\right)z^{p-2} \epsilon  +\frac{3 i a'^3 p^3 \left(p^2-1\right) \left(5 p^2+1\right)z^{p-2}  }{64 \left(4 p^2-1\right)}\epsilon ^3+O(\epsilon^4)\, ,\\
\alpha(r,\theta)&=\text{ln}(\frac{1}{1-r^2})+\frac{a'^2 p}{16 r^2 \left(r^2-1\right)}(4 r^2+4 r^4-p (p+1)^2 r^{2 p}\\&+(p+1)^2 (3 p-4) r^{2 p+2}-(p-1)^2 (3 p+4) r^{2 p+4}+(p-1)^2 p r^{2 p+6})\epsilon ^2\\&+O(\epsilon^4)\, .\\
\end{split}
\label{e10}
\end{equation}

Finally, the area is given by:
\begin{equation}
\cA_{f}=-2 \pi -\frac{1}{2} p \left(p^2-1\right)\pi a'^2 \epsilon ^2+\frac{ p \left(p^2-1\right) \left(23 p^4+p^2\right)\pi a'^4 }{32 \left(4 p^2-1\right)}\epsilon ^4+O(\epsilon^5)\, .
\label{e11}
\end{equation}
Plugging \eqref{e9} into \eqref{e11}, and setting $\epsilon=-iq$, $p=\frac{1}{\nu}$, the area is given by
\begin{equation}
\cA_{f}=-2\pi+\frac{\pi  q^2}{2 \nu  (1-\nu ^2)}-\frac{\left(5 \nu ^2+7\right)\pi  q^4}{32 \nu  \left(4-\nu ^2\right) \left(1-\nu ^2\right)^3}+O(q^6)\, .
\label{e12}
\end{equation}
This result agrees with the one we got from the Mathieu function calculation described in the appendix (see eq.\eqref{ap8}). Notice that, from eq.\eqref{b8}, $q$ is purely imaginary and therefore $\cA_f<-2\pi$ (since $\nu=\frac{1}{n+2}<1$).

\section{Minkowski case}

 It is interesting to consider the case of Minkowski signature since in that case the Wilson loops that we consider approach, in a limit, those used by Alday and Maldacena to compute scattering amplitudes \cite{AM1,AM2}. On the other hand, the methods we employ are similar to the Euclidean case. In Lorentzian signature, in Poincar\'e coordinates $(T,X,Z)$ the metric of \ads{3} is
\beq
 ds^2 = \frac{-dT^2+dX^2+dZ^2}{Z^2}\, .
\label{f1}
\eeq
The Wilson loop is given by a curve
\beq
 x_+(s) = X(s) + T(s) , \ \ \ \ x_-(s)= X(s)-T(s)\, .
\label{f2}
\eeq 
 The light-cone coordinates $x_\pm\in \mathbb{R}$ are introduced for convenience. However it turns out to be even more convenient to use a conformal transformation and define complex coordinates 
\beq
 \hat{x}_{\pm}=\mp i \frac{1\pm ix_{\pm}}{1\mp ix_{\pm}}\, ,
\label{f3}
\eeq
 with the property $|\hat{x}_\pm|=1$. Besides Poincar\'e coordinates it is also useful to use global coordinates $(t,\phi,\rho)$ such that the metric is
\beq
 ds^2 = -\cosh^2\!\rho\, dt^2 + d\rho^2 + \sinh^2\!\rho\, d\phi^2\, .
\label{f4}
\eeq
The relation to Poincar\'e coordinates is better written using embedding coordinates $(X_{-1},X_0,X_1,X_2)$ satisfying $X_{-1}^2 + X_0^2 - X_1^2- X_2^2 = 1$ and related to the previous coordinates by
\beq
Z=\frac{1}{X_{-1}-X_2}\, ,\ 
X=\frac{X_1}{X_{-1}-X_2}\, ,\ 
T=\frac{X_0}{X_{-1}-X_2}\, ,
\label{f5}
\eeq
with the $AdS_3$ boundary located at $Z=0$ while the global coordinates are defined as 
\beq
X_{-1}+iX_0=\cosh{\rho}\, e^{it}\, ,\ 
X_{1}+iX_2=\sinh{\rho}\, e^{i\phi}\, ,
\label{f6}
\eeq 
with the boundary at $\rho\rightarrow\infty$.
 The boundary contour is given by a curve $(t(s),\phi(s))$ and is related to the coordinates $\hat{x}_\pm$ by
\beq
 \hat{x}_\pm = e^{i(t\pm\phi)}\, .
\label{f7}
\eeq
Now we use the method described in \cite{WLMA} for the Euclidean case and generalized in \cite{IK3} for the Minkowski case. In this case the Schroedinger equation associated with the Wilson loop shape reads \cite{IK3}
\beq
\begin{split}
	&-\partial_\theta^2 \chi(\theta) + V_{\lambda}(\theta) \chi(\theta)=0\, ,\\
	V_{\lambda}(\theta)
	&= -\frac{1}{4} + 6\beta_2(\theta)
	+ \frac{1}{\lambda} i f(e^{i\theta}) e^{2i\theta}
	- \lambda i\bar{f}(e^{-i\theta}) e^{-2i\theta}\, .
\end{split}
\label{f8}
\eeq
Given two linearly independent solutions $\chi_{1,2}^{\lambda}$ the shape of the boundary curve is determined as 
\beq
\hat{x}_{\pm} = \frac{\chi_1^{\lambda=\pm1}}{\chi_2^{\lambda=\pm1}}\, .
\label{f9}
\eeq
The solutions $\chi_{1,2}^{\lambda}$ should be chosen such that $|\hat{x}_\pm|=1$.
Equivalently we can chose real solutions $\tilde{\chi}^\lambda_{1,2}$ and define the boundary shape as $x_\pm=\frac{\tilde{\chi}_1^{\lambda=\pm1}}{\tilde{\chi}_2^{\lambda=\pm1}}$.

In the case $f(z)=f_0 z^n$, and after taking $\lambda=e^{i\varphi}$ and defining a new coordinate
\beq
 u=\frac{(n+2)\theta-\varphi}{2}+\frac{\pi}{4}\, ,
\label{f10}
\eeq
we obtain a Mathieu equation with parameters
\beq 
a=\frac{1-24\beta_2}{(n+2)^2},\ q=\frac{4f_0}{(n+2)^2}\,  .
\label{f11}
\eeq
The construction requires the solutions $\chi_{1,2}$ to be anti-periodic. We can then write them in terms of Floquet solutions  
\beq
 \chi_\nu(u+\pi)=e^{i\nu\pi}\chi_\nu(u)\,,\ \ \ \ \nu=\frac{2k+1}{n+2}\ \ \ (k\in Z)\, .
\label{f12}
\eeq
 Noting that the parameters $(a,q)$ in eq.\eqref{f11} are real, the conjugate of a Floquet solution is another Floquet solution. Further,
$u\rightarrow u+\frac{\pi}{2}$ is equivalent to replacing $\lambda \rightarrow -\lambda$, thus, the shape of the Wilson loop can be written as
\beqa
 \hat{x}_+ &=& \frac{\chi_\nu(u)}{(\chi_\nu(u))^*}\, , \\
 \hat{x}_- &=& \frac{(\chi_\nu(u+\frac{\pi}{2}))^*}{\chi_\nu(u+\frac{\pi}{2})}\, ,
\label{f13}
\eeqa
 which evidently satisfy $|\hat{x}_\pm|=1$. Now, for a given value of $\nu$ we find the corresponding value $a_\nu(q)$ and the constant $\beta_2$ in eq.\eqref{f8} follows as
\beq 
\beta_{2}=-\frac{1}{24}\left[1-(n+2)^2 a_\nu(q)\right]\, .
\label{f14}
\eeq 
 The area of the minimal surface ending on the boundary curve can be computed, as before, using integration by parts and results in 
\beqa
 \cA_f &=& -2\pi + 4 \int_D f\bar{f} e^{-2\alpha} d\sigma d\tau \\
       &=& -2\pi + \frac{24\pi\beta_2}{n+2} \\
       &=& -2\pi -\frac{\pi}{n+2} + \pi (n+2) a_\nu(q)\, ,
\label{f15}
\eeqa
 giving again an analytical formula for the area in terms of the eigenvalues of the Mathieu equation. Although the formula is the same as for the Euclidean case, now $\beta_2>0$ and the area is $\cA_f>-2\pi$.

\subsection{Numerics, examples and limits}

 As in the Euclidean case it is useful to plot the resulting surfaces to understand their behavior. To obtain the surface we refer to \cite{IK3} for the derivations and we just present here the summary of steps necessary to obtain the shapes. As a first step we solve the generalized sinh-Gordon equation
\beq
  \partial \bar{\partial} \alpha = e^{2\alpha} - f(z) \bar{f}(\bar{z}) e^{-2\alpha}\, ,
\label{f16}
\eeq
 that again, for $f=f_0\, z^n$ has radial solutions obeying
\beq
 \frac{1}{4}\left[ \partial_r^2 \alpha + \frac{1}{r} \partial_r \alpha \right]=  e^{2\alpha} - f_0^2 r^{2n} e^{-2\alpha}\, .
\label{f17}
\eeq 
 Defining $r_0$ such that $f_0=r_0^{n+2}$ and doing the same rescalings as done to arrive at \eqref{d3} we obtain
 \beq
 \frac{1}{4}\left[ \partial_{\tilde{r}}^2 \tilde{\alpha} + \frac{1}{\tilde{r}} \partial_{\tilde{r}} \tilde{\alpha}\right] =  e^{2\tilde{\alpha}} -  \tilde{r}^{2n} e^{-2\tilde{\alpha}} \, .
 \label{f18}
 \eeq
  We then look for solutions with given value of $\tilde{\alpha}(0)$ and $\left.\partial_{\tilde{r}}\tilde{\alpha}\right|_{\tilde{r}=0}=0$. After that, the linear problem
\beqa
 \partial_{\tilde{r}} \psi_1 &=& \left(\lambda e^{i\theta+\tilde{\alpha}}+\tilde{r}^n e^{-\tilde{\alpha} - i (n+1) \theta}\right)\, \psi_2 \, , \\ 
 \partial_{\tilde{r}} \psi_2 &=&  \left(\frac{1}{\lambda} e^{-i\theta+\tilde{\alpha}}+\tilde{r}^n e^{-\tilde{\alpha}+i(n+1)\theta}\right)\, \psi_1\, ,
\label{f19}
\eeqa 
 needs to be solved for $\lambda=1$ and $\lambda=-1$. Given the initial value $\tilde{\alpha}(0)=\alpha_0$, two linearly independent solutions can be found using, for example the initial conditions, $(\psi^\lambda_1(0)=1, \psi^\lambda_2(0)=-i)$ and $(\tilde{\psi}^\lambda_1(0)=i, \tilde{\psi}^\lambda_2=-1)$. 

\begin{figure}
 	\centerline{\includegraphics[width=\textwidth]{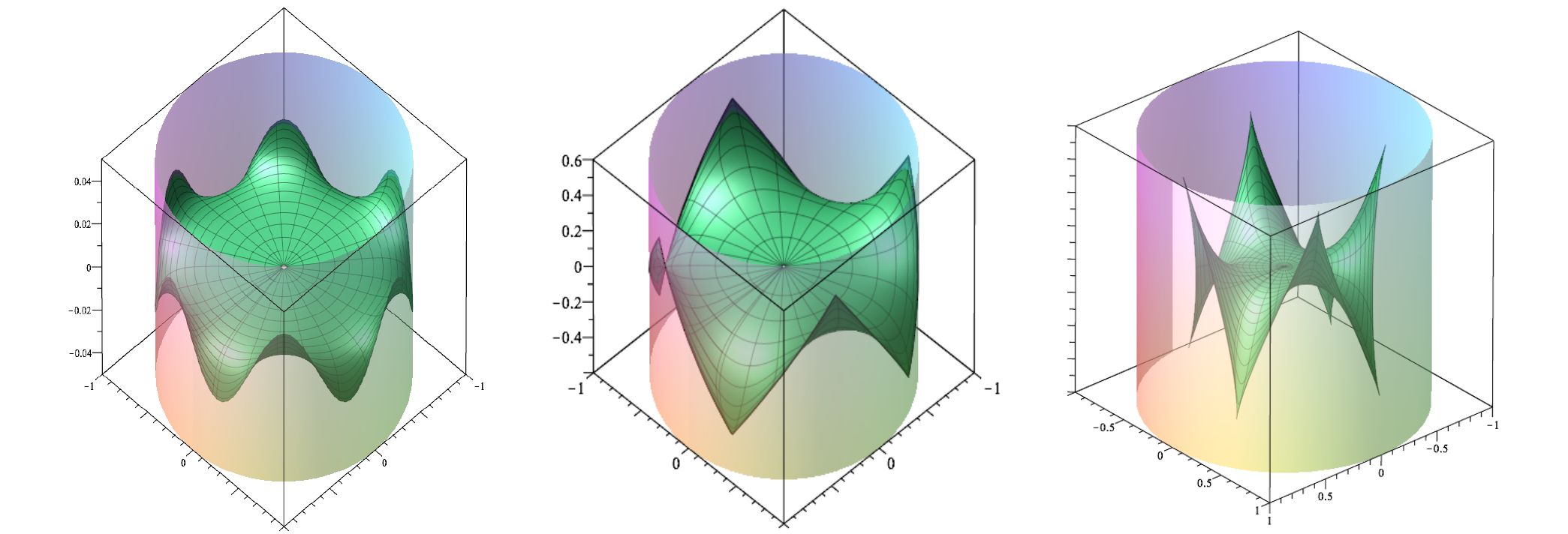}}
 	\caption{Surface in global coordinates for Lorentzian \ads{3} corresponding to the parameters $[n=4, \tilde{\alpha}_0=0]$, $[n=2, \tilde{\alpha}_0=-0.3915941968]$ and $[n=2, \tilde{\alpha}_0=-1]$ respectively.}
 	\label{SurfacesM}
 \end{figure}

 Finally we can reconstruct the shape of the solutions through
\beq
 \mathbb{A}^\lambda = \left(\begin{array}{cc}
 \psi_1^\lambda & \psi_2^\lambda \\
 \tilde{\psi}_1^\lambda & \tilde{\psi}_2^\lambda
 \end{array}\right)\, , \ \ \ \ \ \ \ \ \mathbb{X} = R_M^{-1}\mathbb{A}^{\lambda=1}\left[\mathbb{A}^{\lambda=-1}\right]^{-1}R_M\, ,
\label{f20}
\eeq
 where $R_M$ is a $2\times2$ matrix given by
\beq
 R_M = \frac{1}{\sqrt{2}}\left(\mathbb{I}+i \sigma_1\right), \ \ \ 
 \sigma_1=\left(\begin{array}{cc} 0&1\\ 1&0 \end{array} \right)\, .
\label{f21}
\eeq
Finally, the global coordinates follow as
\beq
 t = \arg(-\mathbb{X}_{11}), \ \ \phi = \arg(\mathbb{X}_{12}), \ \ \tanh\rho=\left|\frac{\mathbb{X}_{12}}{\mathbb{X}_{11}}\right|\, .
\label{f22}
\eeq
 Sweeping the possible values of $\tilde{\alpha}(0)$ we find that, when $\tilde{\alpha}(0)\rightarrow\infty$, the parameter $q$ in the Mathieu equation vanishes and the solution becomes the circular solution $t=0, \phi=\theta$, ($0<\theta<2\pi$). As we lower $\tilde{\alpha}(0)$, the circle starts deforming and takes the shape seen in figure \ref{SurfacesM}, seemingly a regularized version of a succession of light-like cusps. For a certain value of $\tilde{\alpha}(0)$, the parameter $q\rightarrow\infty$ and the Wilson loop becomes a series of light--like segments with the shape of the so called ``regular  polygons'' considered in \cite{AM2}. It has $2(n+2)$ light-like cusps, the particular case $n=0$ has four light-like cusps and was first described in \cite{cusp}, eq.(71). For lower values of $\tilde{\alpha}(0)$, the solution no longer touches the boundary, it still ends in light like-lines but inside AdS. This can be seen in the last figure in figure \ref{SurfacesM}. It will be interesting to understand the physical meaning of such solutions. The particular limits, $q\rightarrow 0 $ and $q\rightarrow\infty$ are studied in the following subsections.

\subsection{Near circular ($q\rightarrow 0$ approximation)}

In the near circular case, we can use a simple generalization of Dekel's procedure from Euclidean to Lorentzian signature to compute the expansion in $q$. It is interesting that it also works in this case. From the point of view of the Mathieu functions, the expansion is the same as the one given in the appendix since it does not depend on the signature of space time. The Wilson loop is now given by two real functions $x_{\pm}(\theta)$. As mentioned, it is convenient to rewrite the contour in terms of 
\begin{equation}
\hat{x}_{\pm}(\theta)=\mp i\frac{1\pm ix_{\pm}(\theta)}{1\mp ix_{\pm}(\theta)}= e^{i(t\pm \phi)}\, ,
\label{f23}
\end{equation}
where $(t,\phi)$ are global coordinates.
Notice that $\hat{x}_{\pm}(\theta)$ have modulus one. In the case of circular Wilson loops, the contour reduces to
\begin{equation}
\hat{x}_{\pm}(\theta)= e^{\pm i \phi}=e^{\pm i \theta}\, .
\label{f24}
\end{equation}
Since the change of variables \eqref{f23} is a global conformal transformation, the Schwarzian derivative with respect to $\theta$ is invariant
\begin{equation}
\{\hat{x}_{\pm}(\theta),\theta\}=\{x_{\pm}(\theta),\theta\}\, .
\label{f25}
\end{equation}
According to the relations \cite{IK3}
\begin{equation}
\{x_{\pm}(\theta),\theta\}=\frac{1}{2} -12\beta_{2}(\theta)\pm 2f e^{2i\theta}\pm 2\bar{f}e^{-2i\theta}\, ,
\label{f26}
\end{equation}
we have
\begin{equation}
\begin{aligned}
&\{\hat{x}_{+},\theta\}+\{\hat{x}_{-},\theta\}=1-24\beta_{2}(\theta)\, ,\\
&\{\hat{x}_{+},\theta\}-\{\hat{x}_{-},\theta\}=4\text{Re}\{e^{2i\theta}f(\theta)\}\, .
\end{aligned}
\label{f27}
\end{equation}
This is the Minkowski version of \eqref{e7a}. Using such relations, we can repeat the procedure of minimal area calculation previously done in the Euclidean case, with the cosh-Gordon equation replaced by the sinh-Gordon equation.

The perturbed contour is parametrized as
\begin{equation}
\hat{x}_{\pm}(\theta)=e^{\pm is(\theta)+i\epsilon j' \sin(p\,s(\theta))}\, ,
\label{f28}
\end{equation} 
where the $i$ in front of the $\epsilon$ is due to the fact that $\hat{x}_{\pm}(\theta)$ has modulus one. 

Repeating the calculation, we have
\begin{equation}
\begin{split}
s(\theta)&=\theta+\frac{j'^2(p+5p^3)\text{sin}(2p\theta)}{8(-1+4p^2)} \epsilon^2+O(\epsilon^4)\, ,\\
f(z)&=-\frac{1}{4}  j' p \left(p^2-1\right)z^{p-2} \epsilon  -\frac{3  j'^3 p^3 \left(p^2-1\right) \left(5 p^2+1\right)z^{p-2}  }{64 \left(4 p^2-1\right)}\epsilon ^3+O(\epsilon^4)\, ,\\
\alpha(r,\theta)&=\text{ln}(\frac{1}{1-r^2})-\frac{j'^2 p}{16 r^2 \left(r^2-1\right)}(4 r^2+4 r^4-p (p+1)^2 r^{2 p}\\&+(p+1)^2 (3 p-4) r^{2 p+2}-(p-1)^2 (3 p+4) r^{2 p+4}+(p-1)^2 p r^{2 p+6})\epsilon ^2\\&+O(\epsilon^4)\, .\\
\end{split}
\label{f29}
\end{equation}
The final result for the area is
\begin{equation}
\cA_{f}=-2 \pi +\frac{1}{2} p \left(p^2-1\right)\pi j'^2 \epsilon ^2+\frac{ p \left(p^2-1\right) \left(23 p^4+p^2\right)\pi j'^4 }{32 \left(4 p^2-1\right)}\epsilon ^4+O(\epsilon^5)\, .
\label{f30}
\end{equation}
The particular contour given by the Mathieu solution is parametrized as
\begin{equation}
\hat{x}_{\pm}(\theta)=e^{\pm is(\theta)+i\xi(s(\theta))}\, ,
\label{f31}
\end{equation}
where
\begin{equation}
\xi(s)=j q \sin \left(\frac{s}{\nu }\right)+q^3 \left(k \sin \left(\frac{s}{\nu }\right)+l \sin \left(\frac{3 s}{\nu }\right)\right)\, .
\label{f32}
\end{equation}
The values of the coefficients are
\begin{equation}
\begin{aligned}
& j=-\frac{\nu }{\nu ^2-1}\, ,\\
& k=-\frac{3 \nu  \left(\nu ^2+5\right)}{16 \left(\nu ^2-4\right) \left(\nu ^2-1\right)^3}\, ,\\
& l=-\frac{\nu  \left(5 \nu ^4+36 \nu ^2-89\right)}{48 \left(\nu ^2-9\right) \left(\nu ^2-4\right) \left(\nu ^2-1\right)^3}\, .
\end{aligned}
\label{f33}
\end{equation}
Similar to the Euclidean case, letting $\epsilon=q$, $p=\frac{1}{\nu}$ and making the replacement
\begin{equation}
j'=j+k\epsilon^2=-\frac{\nu }{\nu ^2-1}-\frac{3 \nu  \left(\nu ^2+5\right) q^2}{16 \left(\nu ^2-4\right) \left(\nu ^2-1\right)^3}\, ,
\label{f34}
\end{equation}
in \eqref{f30}, we have the area
\begin{equation}
\cA_{f}=-2\pi+\frac{\pi  q^2}{2 \nu  \left(1-\nu ^2\right)}-\frac{\left(5 \nu ^2+7\right)\pi  q^4}{32 \nu  \left(4-\nu ^2\right) \left(1-\nu ^2\right)^3}+O(q^5)\, .
\label{f35}
\end{equation}
This result agree with the Mathieu equation calculation and is identical to the result in the Euclidean case. However, now $q$ is real (see eq.\eqref{f11}) and $\cA_f>-2\pi$.

\subsection{Near light-like solution, tight--binding approximation, $q\rightarrow \infty$}

\begin{figure}
  	\centerline{\includegraphics[width=0.5\textwidth]{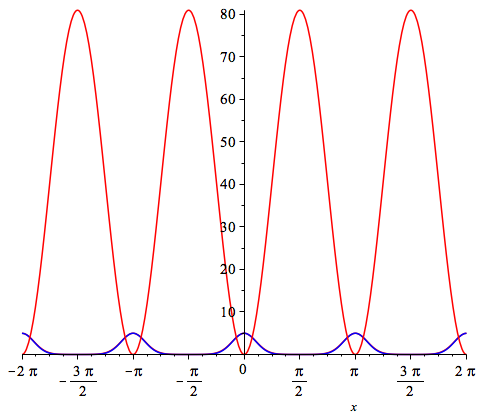}}
  	\caption{Potential (red) and modulus of the Mathieu function (blue) for $q=20.25$ and $n=2$. Clearly the wave-function localizes at the minima.}
  	\label{TBpot}
\end{figure}

The Mathieu equation \eqref{b3} can be written as
\beq
 -\partial_u^2 \chi + V(u) \chi = a \chi, \ \ \ \ V(u) = 2q\cos2u\, ,
\label{f36}
\eeq
which is a standard Schroedinger equation with potential $V(u)$ and energy $a$. From eq.\eqref{b6}, one can see that the cases $\lambda=e^{i\varphi}=\pm1$ correspond to $\varphi=0,\pi$ representing a shift of $0$ or $\pi/2$ in $u$ that simply correspond to taking the potential as $V(u)$ or $-V(u)$. Both cases are similar, let us consider first the case $\lambda=-1$ where the potential is $-V(u)$ which is relevant to determine $x_-(s)$ (whereas $+V(u)$ corresponds to $x_+$). In the limit $q\rightarrow \infty$ such potential becomes a set of $n+2$ separated wells, the tunneling between those is suppressed by a potential barrier of height $q$ (and fixed width). The wave function for the ground state is a superposition of localized wave-functions at the different minima. This is clearly seen in figure \ref{TBpot} where the potential and modulus of the Mathieu function are displayed.
  In this regime, the Mathieu functions can be well approximated by using the tight-binding approximation. It is given by
\beq
  \chi_{\nu}^{TB}(u) = \sum_{j=-\infty}^{\infty} e^{ij\pi\nu}\, \chi_0(u-j\pi) \, .
\label{f36}
\eeq 
 This functions clearly satisfies the periodicity condition $\chi_{\nu}^{TB}(u+\pi) = e^{i\pi\nu} \chi_{\nu}^{TB}(u)$ and obeys (approximately) the wave-equation if $\chi_0(u)$ is chosen appropriately. Here we take $\chi_0$ to be a real polynomial times a Gaussian and choose the polynomial such as to minimize the expectation value of the energy. The result is
\beq
 \chi_0(u) = \left(1+\frac{1}{8} u^2 +\frac{1}{12} \sqrt{q} u^4\right)\, e^{-\sqrt{q} u^2}\, ,
\label{f37}
\eeq
with energy given by
\beq
 a_\nu(q) = -2 q + 2\sqrt{q} - \frac{1}{4} - \frac{1}{32}\frac{1}{\sqrt{q}} - \frac{3}{256} \frac{1}{q} + \cO\left(\frac{1}{q^{3/2}}\right)\, .
\label{f38}
\eeq
 Notice that in this approximation $a_\nu(q)$ is independent of $\nu$, namely around the given value of $a_\nu(q)$ there is an exponentially small window where $\nu$ changes from 0 to 1. An improved result is obtained by replacing the wave function $\chi_0$ away from $u=0$ by a WKB approximation:
\beq
 \chi_0^{WKB} = \frac{C}{\kappa(u)}\, e^{\int^u \kappa(u')du'}  , \ \ \ \ \kappa(u)=\sqrt{-a(q)-2q\cos 2u}\, ,
\label{f39}
\eeq 
 where the constant $C$ is chosen to match the previous approximation $\chi_0(u)$ at some intermediate point. 
 
 Since the shape is given by
\beq
 \hat{x}_- = \frac{\left(\chi^{TB}_\nu\right)^*}{\chi^{TB}_\nu}\, ,
\label{f40}
\eeq
 and the function $\chi_0(u)$ is real, the phase of $\chi^{TB}_\nu$ changes abruptly 
when $u$ crosses a maximum of the potential. For example at $u=\frac{\pi}{2}$ in fig. \ref{TBpot}, the wave function localized around $u=0$ is replaced by the function localized at $u=\pi$ that is multiplied by a different phase in eq.\eqref{f36}. This abrupt changes correspond to the straight lines along $x_-$. When we consider $\lambda=+1$ the potential is inverted and therefore the minima of the potential correspond to straight lines along $x_+$. The cusps are associated to the intermediate regions between maxima and minima. In those regions, the phase of $\chi_\nu$ remains constant implying that $\hat{x}_+$, $\hat{x}_-$ are also constant and equal to their value at the cusp.

It seems possible to define a general procedure to expand around the light-like Wilson loop similar to what can be done around the circle. We leave that for future work. We conclude this part by computing the area in this approximation using eqs.\eqref{f15} and \eqref{f38}:
\beq
\cA_f = -2\pi- \frac{\pi}{n+2}+\pi(n+2) \left[-2 q + 2\sqrt{q} - \frac{1}{4} - \frac{1}{32}\frac{1}{\sqrt{q}} - \frac{3}{256} \frac{1}{q} + \cO\left(\frac{1}{q^{3/2}}\right) \right]\, .
\label{f41}
\eeq
When $q\rightarrow\infty$ there is a divergence proportional to the number of cusps as we expect from the cusp anomaly computations \cite{cusp, AM1,AMSV}.

\section{Conclusions}

 One of the most important observables in gauge theories is the Wilson loop. The AdS/CFT correspondence opened up the possibility of studying such operator at strong 't Hooft coupling by using minimal area surfaces. In this paper we consider a previously proposed method to find such surfaces based on the properties of a Schroedinger equation with a potential given by the Schwarzian derivative of the shape of the Wilson loop. It allows us to find new interesting Wilson loops dual to surfaces whose area can be computed analytically in terms of eigenvalues of the Mathieu equation. Those Wilson loops have interesting properties since they can be seen as deformed or regularized versions of previously known solutions: the regular polygons with light-like cusps and also of the multiple wound circular Wilson loop. It also allowed us to check analytically Dekel's method to expand around the circular solutions. Finally, in the near light--like case, we found that the potential becomes a series of separated wells each one associated with a light-like segment. This opens up the possibility of studying systematically perturbations around such Wilson loop, a subject that we leave for future work. Summarizing, we believe that the idea of studying Wilson loops through its associated Schroedinger equation is a productive one that uses the integrability properties of the system and that presumably will lead to further insight into this important operator. 

\section{Acknowledgments}

 We are very grateful to A. Dekel, S. Komatsu, J. Toledo and P. Vieira for useful comments and discussions on this topic. 
 This work was supported in part by DOE through grant DE-SC0007884. 
 M.K. is very grateful to Perimeter Institute for Theoretical Physics for hospitality while this work was being completed. 

\appendix

\section{Expansion of Mathieu functions}

 To check the perturbative expansion it is necessary to expand the Mathieu functions for small values of the parameter $q$ appearing in eq.(\ref{b3}). More precisely, given $q\ll1$ and $\nu$, the eigenvalue $a_\nu(q)$ and eigenfunctions $\chi_\nu$ are expanded as
\beqa
 a_\nu(q) &=& \nu^2 + \sum_{j=1}^\infty a_j q^j\, , \\
 \chi_\nu(u) &=& e^{i\nu u} \sum_{n=-\infty}^{\infty} c_n(q) e^{2inu} \, ,
 \label{ap1}
\eeqa
with the Fourier coefficients $c_n$ expanded as 
\beq
  c_n = \sum_{j=1}^{\infty} c_n^{(j)}\, q^j, \ \ \ \ (n\neq 0)\, ,
  \label{ap2}
\eeq
and we take $c_0=1$ as a normalization condition. Replacing in the equation a simple recursion relation is obtained that allows an easy computation of the coefficients at any desired order:  
\beqa
a_j &=& c_{-1}^{(j-1)} + c_1^{(j-1)}\, , \\
c_n^{(j)} &=& -\frac{1}{4n(n+\nu)}\left[c^{(j-1)}_{n-1}+c_{n+1}^{(j-1)} - \sum_{p=1}^{j-1} a_p c_n^{(j-p)}\right]\, ,
\label{ap3}
\eeqa
where $n\neq 0$ and $j\ge 1$.
These results can then be used in eqns.(\ref{b13}) and (\ref{c3}) to expand the shape and the area in powers of $q$. The expansion of the shape is, schematically:
\beq
 X = \frac{\chi_\nu(u)}{\chi_{\nu}(-u)} = e^{2i\nu u+ \chi_1 q \sin 2u+ \chi_2 q^2 \sin 4u + \chi_{3a} q^3 \sin 2u + \chi_{3b} q^3 \sin 6u + \cO(q^4) }\, ,
\label{ap4}
\eeq
where $\chi_l$ are constant coefficients. By defining a new variable 
\beq
s = 2 \nu u + \chi_2 q^2 \sin 4u + \cO(q^4)\, ,
\label{ap5}
\eeq
the shape can be written as
\beq
 X = e^{is + \xi(s)}\, ,
\label{ap6}
\eeq
where
\beqa
 s&=& 2 \nu u -\frac{\nu(\nu^2+5)}{8(\nu^2-1)^2(\nu^2-4)} q^2 \sin 4u + \cO(q^4)\, , \\
 \xi(s) &=&\frac{iq\nu}{1-\nu^2} \sin\frac{s}{\nu} -\frac{3}{16}\frac{iq^3\nu(\nu^2+5)}{(\nu^2-1)^3(\nu^2-4)}  \sin\frac{s}{\nu} \\
  && - \frac{iq^3\nu(5\nu^4+36\nu^2-89)}{48(\nu^2-1)^3(\nu^2-4)(\nu^2-9)}  \sin\frac{3s}{\nu} +\cO(q^5)\, .
\label{ap7}
\eeqa 
Notice that $\xi(s)$ is real since $q$ is purely imaginary (for the Euclidean case). 
Finally, the area follows from eq.(\ref{c3}) as:
\beqa
\cA_f &=& - 2\pi + \pi \nu - \frac{\pi}{\nu} a_\nu(q) \\
      &=& -2\pi + \half \frac{\pi}{\nu(1-\nu^2)}\, q^2 - \frac{1}{32}\frac{\pi(5\nu^2+7)}{\nu(4-\nu^2)(1-\nu^2)^3}\, q^4\, , +\cO(q^6)
\label{ap8}
\eeqa
where we used $\nu=\frac{1}{n+2}$. This result for the area can also be obtained, as done in the main text, by using the shape of the Wilson loop in Dekel's method. The result agrees perfectly providing a useful check.


\begin{thebibliography}{99}        


    
\bibitem{malda}
J.~Maldacena,
``The large $N$ limit of superconformal field theories and supergravity,''
Adv.\ Theor.\ Math.\ Phys.\  {\bf 2}, 231 (1998)
[Int.\ J.\ Theor.\ Phys.\  {\bf 38}, 1113 (1998)],
{\tt hep-th/9711200}, \\
S.~S.~Gubser, I.~R.~Klebanov and A.~M.~Polyakov,
``Gauge theory correlators from non-critical string theory,''
Phys.\ Lett.\ B {\bf 428}, 105 (1998)
[arXiv:hep-th/9802109], \\
E.~Witten,
``Anti-de Sitter space and holography,''
Adv.\ Theor.\ Math.\ Phys.\  {\bf 2}, 253 (1998)
[arXiv:hep-th/9802150]. 

\bibitem{largeN}
G.'t Hooft, Nucl. Phys. {\bf B72} (1974) 461, 
G.'t Hooft, Nucl. Phys. {\bf B75} (1974) 461. 



\bibitem{MRY}
  J.~M.~Maldacena,
  ``Wilson loops in large N field theories,''
  Phys.\ Rev.\ Lett.\  {\bf 80}, 4859 (1998)
  [arXiv:hep-th/9803002], \\
  S.~J.~Rey and J.~T.~Yee,
  ``Macroscopic strings as heavy quarks in large N gauge theory and anti-de
  Sitter supergravity,''
  Eur.\ Phys.\ J.\  C {\bf 22}, 379 (2001)
  [arXiv:hep-th/9803001].

\bibitem{cWL}
  D.~E.~Berenstein, R.~Corrado, W.~Fischler and J.~M.~Maldacena,
  ``The Operator product expansion for Wilson loops and surfaces in the large N
  limit,''
  Phys.\ Rev.\  D {\bf 59}, 105023 (1999)
  [arXiv:hep-th/9809188], \\
  D.~J.~Gross and H.~Ooguri,
   ``Aspects of large N gauge theory dynamics as seen by string theory,''
   Phys.\ Rev.\  D {\bf 58}, 106002 (1998)
   [arXiv:hep-th/9805129], \\
  J.~K.~Erickson, G.~W.~Semenoff and K.~Zarembo,
  ``Wilson loops in N = 4 supersymmetric Yang-Mills theory,''
  Nucl.\ Phys.\  B {\bf 582}, 155 (2000)
  [arXiv:hep-th/0003055], \\
  N.~Drukker and D.~J.~Gross,
  ``An exact prediction of N = 4 SUSYM theory for string theory,''
  J.\ Math.\ Phys.\  {\bf 42}, 2896 (2001)
  [arXiv:hep-th/0010274], \\
  V.~Pestun,
  ``Localization of gauge theory on a four-sphere and supersymmetric Wilson
  loops,''
  arXiv:0712.2824 [hep-th], \\
    M.~Kruczenski and A.~Tirziu,
    ``Matching the circular Wilson loop with dual open string solution at 1-loop
    in strong coupling,''
    JHEP {\bf 0805}, 064 (2008)
    [arXiv:0803.0315 [hep-th]], \\
    A.~Faraggi and L.~A.~P.~Zayas,
    ``The Spectrum of Excitations of Holographic Wilson Loops,''
    arXiv:1101.5145 [hep-th], \\
 E.~I.~Buchbinder and A.~A.~Tseytlin,
  ``The 1/N correction in the D3-brane description of circular Wilson loop at strong coupling,''
  arXiv:1404.4952 [hep-th],\\
  V.~Forini, V.~Giangreco, M.~Puletti, L.~Griguolo, D.~Seminara and E.~Vescovi,
  ``Precision calculation of 1/4-BPS Wilson loops in AdS$_5\times S^5$,''
  JHEP {\bf 1602}, 105 (2016)
  doi:10.1007/JHEP02(2016)105
  [arXiv:1512.00841 [hep-th]],\\
    A.~Faraggi, L.~A.~Pando Zayas, G.~A.~Silva and D.~Trancanelli,
    ``Toward precision holography with supersymmetric Wilson loops,''
    arXiv:1601.04708 [hep-th].


\bibitem{WLref}
N.~Drukker, D.~J.~Gross and H.~Ooguri,
  ``Wilson loops and minimal surfaces,''
  Phys.\ Rev.\  D {\bf 60}, 125006 (1999)
  [arXiv:hep-th/9904191], \\
  N.~Drukker, S.~Giombi, R.~Ricci and D.~Trancanelli,
  ``Supersymmetric Wilson loops on S**3,''
  JHEP {\bf 0805}, 017 (2008)
  [arXiv:0711.3226 [hep-th]], \\
  N.~Drukker and B.~Fiol,
  ``On the integrability of Wilson loops in AdS(5) x S**5: Some periodic
  ansatze,''
  JHEP {\bf 0601}, 056 (2006)
  [arXiv:hep-th/0506058], \\
    K.~Zarembo,
    ``Supersymmetric Wilson loops,''
    Nucl.\ Phys.\  B {\bf 643}, 157 (2002)
    [arXiv:hep-th/0205160],
     N.~Drukker, S.~Giombi, R.~Ricci and D.~Trancanelli,
                         ``Supersymmetric Wilson loops on S**3,''
                         JHEP {\bf 0805}, 017 (2008)
                         [arXiv:0711.3226 [hep-th]], \\
                            N.~Drukker,
                            ``1/4 BPS circular loops, unstable world-sheet instantons and the matrix model,''
                            JHEP {\bf 0609}, 004 (2006)
                            [hep-th/0605151].

 
\bibitem{WLint}
               B.~Fiol and G.~Torrents,
               ``Exact results for Wilson loops in arbitrary representations,''
               arXiv:1311.2058 [hep-th], \\
                  D.~Muller, H.~Munkler, J.~Plefka, J.~Pollok and K.~Zarembo,
                  ``Yangian Symmetry of smooth Wilson Loops in N=4 super Yang-Mills Theory,''
                  arXiv:1309.1676 [hep-th], \\
                      H.~Munkler and J.~Pollok,
                      ``Minimal Surfaces of the $AdS_5\times S^5$ Superstring and the Symmetries of Super Wilson Loops at Strong Coupling,''
                      arXiv:1503.07553 [hep-th], \\
                  S.~Ryang,
                  ``Algebraic Curves for Long Folded and Circular Winding Strings in AdS5xS5,''
                  JHEP {\bf 1302}, 107 (2013)
                  [arXiv:1212.6109 [hep-th]], \\
                                  A.~Dekel,
                                  ``Algebraic Curves for Factorized String Solutions,''
                                  JHEP {\bf 1304}, 119 (2013)
                                  [arXiv:1302.0555 [hep-th]], \\
                                  A.~Dekel and T.~Klose,
                                    ``Correlation Function of Circular Wilson Loops at Strong Coupling,''
                                    JHEP {\bf 1311}, 117 (2013)
                                    [arXiv:1309.3203 [hep-th]], \\
                  A.~Irrgang and M.~Kruczenski,
                                                       ``Double-helix Wilson loops: Case of two angular momenta,''
                                                       JHEP {\bf 0912}, 014 (2009)
                                                       [arXiv:0908.3020 [hep-th]], \\
                  V.~Forini, V.~G.~M.~Puletti and O.~Ohlsson Sax,
                  ``Generalized cusp in $AdS_4 \times CP^3$ and more one-loop results from semiclassical strings,''
                  J.\ Phys.\ A {\bf 46}, 115402 (2013)
                  [arXiv:1204.3302 [hep-th]], \\
                    B.~A.~Burrington and L.~A.~Pando Zayas,
                    ``Phase transitions in Wilson loop correlator from integrability in global AdS,''
                    Int.\ J.\ Mod.\ Phys.\ A {\bf 27}, 1250001 (2012)
                    [arXiv:1012.1525 [hep-th]], \\
               G.~Papathanasiou,
               ``Pohlmeyer reduction and Darboux transformations in Euclidean worldsheet $AdS_3$,''
               JHEP {\bf 1208}, 105 (2012)
               [arXiv:1203.3460 [hep-th]], \\
               N.~Drukker and V.~Forini,
               ``Generalized quark-antiquark potential at weak and strong coupling,''
               JHEP {\bf 1106}, 131 (2011)
               [arXiv:1105.5144 [hep-th]], \\
               B.~A.~Burrington,
               ``General Leznov-Savelev solutions for Pohlmeyer reduced AdS$_5$ minimal surfaces,''
               JHEP {\bf 1109}, 002 (2011)
               [arXiv:1105.3227 [hep-th]], \\
               L.~F.~Alday and A.~A.~Tseytlin,
               ``On strong-coupling correlation functions of circular Wilson loops and local operators,''
               J.\ Phys.\ A {\bf 44}, 395401 (2011)
               [arXiv:1105.1537 [hep-th]], \\
               C.~Kalousios and D.~Young,
               ``Dressed Wilson Loops on $S^2$,''
               Phys.\ Lett.\ B {\bf 702}, 299 (2011)
               [arXiv:1104.3746 [hep-th]],\\
                R.~Ishizeki, M.~Kruczenski and A.~Tirziu,
                 ``New open string solutions in AdS(5),''
                 Phys.\ Rev.\ D {\bf 77}, 126018 (2008)
                 [arXiv:0804.3438 [hep-th]], \\
                  R.~A.~Janik and P.~Laskos-Grabowski,
                   ``Surprises in the AdS algebraic curve constructions: Wilson loops and correlation functions,''
                   Nucl.\ Phys.\ B {\bf 861}, 361 (2012)
                   [arXiv:1203.4246 [hep-th]].
            
 


\bibitem{SY} 
  G.~W.~Semenoff and D.~Young,
  ``Wavy Wilson line and AdS / CFT,''
  Int.\ J.\ Mod.\ Phys.\ A {\bf 20}, 2833 (2005)
  [hep-th/0405288].

\bibitem{CHMS} 
  D.~Correa, J.~Henn, J.~Maldacena and A.~Sever,
  ``An exact formula for the radiation of a moving quark in N=4 super Yang Mills,''
  JHEP {\bf 1206}, 048 (2012)
  [arXiv:1202.4455 [hep-th]].
 
 \bibitem{Cagnazzo} 
   A.~Cagnazzo,
   ``Integrability and Wilson loops: the wavy line contour,''
   arXiv:1312.6891 [hep-th].

\bibitem{Dekel} 
  A.~Dekel,
  ``Wilson Loops and Minimal Surfaces Beyond the Wavy Approximation,''
  JHEP {\bf 1503}, 085 (2015)
  [arXiv:1501.04202 [hep-th]].
   
\bibitem{IKZ} 
  R.~Ishizeki, M.~Kruczenski and S.~Ziama,
  ``Notes on Euclidean Wilson loops and Riemann Theta functions,''
  Phys.\ Rev.\ D {\bf 85}, 106004 (2012)
  [arXiv:1104.3567 [hep-th]].

\bibitem{KZ} 
  M.~Kruczenski and S.~Ziama,
  ``Wilson loops and Riemann theta functions II,''
  JHEP {\bf 1405}, 037 (2014)
  [arXiv:1311.4950 [hep-th]].

\bibitem{IK} 
  A.~Irrgang and M.~Kruczenski,
  ``Rotating Wilson loops and open strings in AdS3,''
  J.\ Phys.\ A {\bf 46}, 075401 (2013)
  [arXiv:1210.2298 [hep-th]].

\bibitem{BB}
M.~Babich and A.~Bobenko,
``Willmore Tori with umbilic lines and minimal surfaces in hyperbolic space'',
 Duke Mathematical Journal {\bf 72, No. 1}, 151 (1993).

\bibitem{BBook}
E.~D.~Belokolos, A.~I.~Bobenko,V.~Z.~Enol'skii, A.~R.~Its, V.~B.~Matveev,
``Algebro-Geometric Approach to Nonlinear Integrable Equations,''
Springer-Verlag series in Non-linear Dynamics,
Springer-Verlag Berlin Heidelberg NewYork (1994).

\bibitem{Toledo} 
  J.~C.~Toledo,
  ``Smooth Wilson loops from the continuum limit of null polygons,''
  arXiv:1410.5896 [hep-th].

\bibitem{cusp}
M.~Kruczenski,
  ``A note on twist two operators in N = 4 SYM and Wilson loops in Minkowski
signature,''
  JHEP {\bf 0212}, 024 (2002)
  [arXiv:hep-th/0210115].


\bibitem{scatampl}
See e.g. \\
 L.~F.~Alday and J.~M.~Maldacena,
  ``Gluon scattering amplitudes at strong coupling,''
  JHEP {\bf 0706}, 064 (2007)
  [arXiv:0705.0303 [hep-th]], \\ 
  L.~F.~Alday and J.~Maldacena,
  ``Comments on gluon scattering amplitudes via AdS/CFT,''
  JHEP {\bf 0711}, 068 (2007)
  [arXiv:0710.1060 [hep-th]], \\
  J.~Maldacena and A.~Zhiboedov,
  ``Form factors at strong coupling via a Y-system,''
  JHEP {\bf 1011}, 104 (2010)
  [arXiv:1009.1139 [hep-th]], \\
  L.~F.~Alday, B.~Eden, G.~P.~Korchemsky, J.~Maldacena and E.~Sokatchev,
  ``From correlation functions to Wilson loops,''
  arXiv:1007.3243 [hep-th], \\
  L.~F.~Alday, D.~Gaiotto, J.~Maldacena, A.~Sever and P.~Vieira,
  ``An Operator Product Expansion for Polygonal null Wilson Loops,''
  arXiv:1006.2788 [hep-th],  \\
  H.~Dorn, N.~Drukker, G.~Jorjadze and C.~Kalousios,
        ``Space-like minimal surfaces in AdS x S,''
        JHEP {\bf 1004}, 004 (2010)
        [arXiv:0912.3829 [hep-th]].

\bibitem{AMSV}
  L.~F.~Alday, J.~Maldacena, A.~Sever and P.~Vieira,
  ``Y-system for Scattering Amplitudes,''
  J.\ Phys.\ A  {\bf 43}, 485401 (2010)
  [arXiv:1002.2459 [hep-th]].

\bibitem{ClosedStrings}
  A.~Jevicki and K.~Jin,
  ``Moduli Dynamics of AdS(3) Strings,''
  JHEP {\bf 0906}, 064 (2009)
  [arXiv:0903.3389 [hep-th]],\\
  A.~Jevicki, K.~Jin, C.~Kalousios and A.~Volovich,
           ``Generating AdS String Solutions,''  
           JHEP {\bf 0803}, 032 (2008)
           [arXiv:0712.1193 [hep-th]], \\   
            M.~Kruczenski,
             ``Spin chains and string theory,''
             Phys.\ Rev.\ Lett.\  {\bf 93}, 161602 (2004)
             [arXiv:hep-th/0311203], \\       
M.~Kruczenski,
  ``Spiky strings and single trace operators in gauge theories,''
  JHEP {\bf 0508}, 014 (2005)
  [arXiv:hep-th/0410226], \\
 N.~Dorey and B.~Vicedo,
  ``On the dynamics of finite-gap solutions in classical string theory,''
  JHEP {\bf 0607}, 014 (2006)
  [arXiv:hep-th/0601194], \\
  K.~Sakai and Y.~Satoh,
  ``Constant mean curvature surfaces in \ads{3},''
  JHEP {\bf 1003}, 077 (2010)
  [arXiv:1001.1553 [hep-th]], \\
H.~J.~De Vega and N.~G.~Sanchez,
          ``Exact integrability of strings in D-Dimensional De Sitter space-time,''
          Phys.\ Rev.\ D {\bf 47}, 3394 (1993), \\
             A.~L.~Larsen and N.~G.~Sanchez,
             ``Sinh-Gordon, cosh-Gordon and Liouville equations for strings and multistrings in constant curvature space-times,''  
             Phys.\ Rev.\ D {\bf 54}, 2801 (1996)  [hep-th/9603049], \\
K.~Zarembo,
                               ``Wilson loop correlator in the AdS / CFT correspondence,''
                               Phys.\ Lett.\ B {\bf 459}, 527 (1999)
                               [hep-th/9904149], \\
 N.~Drukker and B.~Fiol,
           ``On the integrability of Wilson loops in AdS(5) x S**5: Some periodic
           ansatze,''
           JHEP {\bf 0601}, 056 (2006)
           [arXiv:hep-th/0506058].

 \bibitem{Pohlmeyer}
 K. Pohlmeyer, 
 ``Integral Hamiltonian systems and interactions through quadratic constraints,'' 
Commun. Math. Phys. 46, 207 (1976). 
           

\bibitem{WLMA} 
  M.~Kruczenski,
  ``Wilson loops and minimal area surfaces in hyperbolic space,''
  JHEP {\bf 1411}, 065 (2014)
  [arXiv:1406.4945 [hep-th]].

\bibitem{Mathieu}
See \eg\ 
Gradshteyn and Ryzhik, ``Table of Integrals Series and Products'',
A. Jeffrey, D. Zwillinger Editors, Academic Press, San Diego, CA, USA, London, UK, (2000), \\
McLachlan, N.W. , ``Theory and Applications of Mathieu functions'', Clarendon Press Oxford (1947), \\ 
and for more general periodic potentials: \\
Wilhelm Magnus, Stanley Winkler , 
 "Hill's Equation", Dover Books on Mathematics, Dover Publications (2004).
      

\bibitem{AM2} 
  L.~F.~Alday and J.~Maldacena,
  ``Null polygonal Wilson loops and minimal surfaces in Anti-de-Sitter space,''
  JHEP {\bf 0911}, 082 (2009)
  doi:10.1088/1126-6708/2009/11/082
  [arXiv:0904.0663 [hep-th]].
\bibitem{Novokshenov} 
      V.~Y.~Novokshenov,
      ``Minimal surfaces in the hyperbolic space and radial symmetric solutions of the Cosh-Laplace equation,''
INS-251-CLARKSON.
      
\bibitem{IK3} 
        A.~Irrgang and M.~Kruczenski,
        ``Euclidean Wilson loops and minimal area surfaces in lorentzian AdS$_{3}$,''
        JHEP {\bf 1512}, 083 (2015)
        doi:10.1007/JHEP12(2015)083
        [arXiv:1507.02787 [hep-th]].

\bibitem{AM1} 
    L.~F.~Alday and J.~M.~Maldacena,
    ``Gluon scattering amplitudes at strong coupling,''
    JHEP {\bf 0706}, 064 (2007)
    doi:10.1088/1126-6708/2007/06/064
    [arXiv:0705.0303 [hep-th]].




\end{thebibliography}
\end{document}